\documentclass[a4paper,fleqn,usenatbib]{mnras}
\usepackage{amssymb,amsmath} % for astro-ph submission
\usepackage{txfonts} % for astro-ph submission
\usepackage[T1]{fontenc}
\usepackage{ae,aecompl}

\usepackage{graphicx}	% Including figure files
\usepackage{natbib}

\title[Discovery of a magnetic field in $\rho$\,Pup]{Discovery of a magnetic field in the $\delta$\,Scuti F2m star $\rho$\,Pup}

\author[C. Neiner et al.]{
C. Neiner,$^{1}$\thanks{E-mail: coralie.neiner@obspm.fr}
G.~A. Wade,$^{2}$
and J. Sikora$^{2}$
\\
$^{1}$LESIA, Observatoire de Paris, PSL Research University, CNRS, Sorbonne Universit\'es, UPMC Univ. Paris 06, Univ. Paris\\
Diderot, Sorbonne Paris Cit\'e, 5 place Jules Janssen, 92195 Meudon, France\\
$^{2}$Department of Physics, Royal Military College of Canada, PO Box 17000 Kingston, ON K7K 7B4, Canada
}

\date{Accepted XXX. Received YYY; in original form ZZZ}
\pubyear{2016}

\begin{document}
\label{firstpage}
\pagerange{\pageref{firstpage}--\pageref{lastpage}}
\maketitle

\begin{abstract}
$\rho$\,Pup is a $\delta$\,Scuti F2 pulsator, known to host a main radial mode
as well as non-radial pulsations, with chemical peculiarities typical of evolved
Am stars. We present high-precision spectropolarimetric observations of this
star, obtained with ESPaDOnS at the Canada France Hawaii Telescope (CFHT) in the
frame of the BRITE spectropolarimetric survey. A magnetic field is clearly
detected in $\rho$\,Pup, with a longitudinal field strength below 1 G. This
makes $\rho$\,Pup the second known magnetic $\delta$\,Scuti discovered, after
HD\,188774, and a possible cool evolved counterpart of the recently discovered
ultra-weakly magnetic Am family.
\end{abstract}


%%%%%%%%%%%%%%%%%%%%%%%%%%%%%%%%%%%%%%%%%%%%%%%%%%

%%%%%%%%%%%%%%%%% BODY OF PAPER %%%%%%%%%%%%%%%%%%

\section{Introduction}

$\delta$\,Scuti variables are A and F stars pulsating with low radial order
pressure modes and mixed modes driven by the $\kappa$\,mechanism operating in
the \ion{He}{ii} ionization zone \citep{pamyatnykh2000} and by the turbulent pressure in the hydrogen ionisation layer \citep{antoci2014,xiong2016}. Their pulsation periods range from approximately
15 min to 8 h. $\delta$\,Scuti stars can be on the pre-main sequence (PMS), main
sequence (MS), or start of the post-MS. They have masses from 1.5 to 2.5
M$_\odot$ and effective temperatures between 6700 and 8000 K. Between 6900 and
7400 K, $\delta$\,Scuti stars can also host gravity modes of $\gamma$\,Dor type.
These pulsations are driven by convective blocking near the base of their
convective envelopes \citep{guzik2000,dupret2004}. The stars showing both types
of pulsations at the same time are called hybrid stars. 

$\rho$\,Pup (HD\,67523) is a bright (V=2.81) and long-known $\delta$\,Scuti star
\citep[e.g.][]{cousins1951,eggen1956,ponsen1963,mathias1997,dall2002}. Its main
pulsation frequency $f=7.0972$ c~d$^{-1}$ has been identified as the radial
fundamental mode, and it also hosts at least one non-radial mode with $f=7.9$
c~d$^{-1}$ \citep{mathias1997,tkachenko2013,antoci2013,nardetto2014}. In
addition, a search for solar-like oscillations was performed for this star
thanks to a spectroscopic multisite campaign, but no such pulsations were
detected \citep{antoci2013}. 

The stellar parameters of $\rho$\,Pup have been determined to be within $T_{\rm
eff}=6500-6850$ K and $\log g=3.3-3.7$ dex, depending on the authors
\citep[e.g.][]{kurtz1976,prugniel2011,nardetto2014}. According to Hipparcos
measurements, $\rho$\,Pup is located at $19.48 \pm 0.06$ pc \citep{leeuwen2007}.
Its radius is $R=3.52 \pm 0.07$ R$_\odot$ \citep{antoci2013} and its absolute
angular diameter varies by  11 $\mu$as (or 0.7\%) due to the pulsations
\citep{nardetto2014}.

Moreover, $\rho$\,Pup is an evolved F2 star, showing enhanced metal lines
\citep{kurtz1976}. \cite{kurtz1976} and \cite{gray1989} suggested that
$\rho$\,Pup is a cool, evolved Am star, and \cite{yushchenko2015} confirmed that
the abundance pattern of $\rho$\,Pup is similar to Am stars. In fact, stars
showing such types of peculiar spectra have been named $\rho$\,Puppis stars by
\cite{gray1989}. Other authors also called them $\delta$\,Delphini stars,
although this class contained varied types of spectra and the name is therefore
deprecated now.

Finally, no companion was detected for this star \citep{nardetto2014}.

\section{Spectropolarimetry}

\subsection{Observations}

The BRITE spectropolarimetric survey aims at acquiring very high signal-to-noise
(S/N) Stokes V spectra of all stars brighter than V=4 with high-resolution
spectropolarimeters \citep{neiner2016_innsbruck}. This program provides
ground-based support for the BRITE (BRIght Target Explorer) constellation of
nano-satellites, which performs photometric observations of bright stars, in
particular for seismology \citep{weiss2014}.

In this frame, we observed $\rho$\,Pup with the ESPaDOnS spectropolarimeter at
the Canada France Hawaii Telescope (CFHT) in Hawaii. ESPaDOnS covers a spectral
range from about 375 to 1050 nm, with a resolving power of $\sim$68000, spread
on 40 echelle orders.

We used the circular polarisation mode to measure the Stokes V spectrum together
with the intensity spectrum (Stokes I). Each Stokes V sequence consists of the
combination of 4 sub-exposures obtained with the polarimetry half-wave Fresnel
rhombs set at various angles. The sub-exposures are also destructively combined
to produce a null polarisation (N) spectrum to check for pollution by, e.g.,
variable observing conditions, instrumental effects, or non-magnetic stellar
effects such as pulsations. In addition, successive Stokes V sequences have been
acquired to increase the total S/N ratio of a magnetic measurement.

$\rho$\,Pup was observed a first time on February 10, 2014, for 2 successive
sequences of 4$\times$30 s each, and a second time on October 31, 2015, for 8
successive sequences of 4$\times$25 s (see Table~\ref{tableobs}). Following the
ephemeris provided by \cite{nardetto2014}, this corresponds to pulsation phases
0.87 and 0.49, respectively. The duration of each Stokes V sequence, including
the detector readout time ($\sim$38 s), is either 234 or 214 s, which is less
than 1/50$^{th}$ of the main pulsation period and corresponds to a radial
velocity change below 0.2 km~s$^{-1}$ (i.e. well below the 1.8 km~s$^{-1}$
resolution of the data). This insures that the spectropolarimetric observations
are not influenced by line profile variations or radial velocity shifts due to
pulsations.

\begin{table}
\caption{Journal of ESPaDOnS observations of $\rho$\,Pup indicating 
the Heliocentric Julian Date at the middle of the observations (mid-HJD -
2450000), the exposure time, and the signal-to-noise ratio of the
(co-added) spectrum at $\sim$500 nm.}
\begin{tabular}{lllllr}
\hline
\# & Date & mid-HJD & T$_{\rm exp}$ & S/N \\
 & & -2450000 & s &  \\
\hline
1 & Feb 10, 2014  & 6698.86667 & 2x4x30  & 1416 \\
2 & Oct 31, 2015  & 7327.14429 & 8x4x25  & 3528 \\
\hline
\end{tabular}
\label{tableobs}
\end{table}

The data were reduced with the Upena pipeline feeding the Libre-Esprit reduction
package \citep{donati1999} available at CFHT. This included standard bias
removal, flat-fielding, and wavelength calibration. The Stokes I spectra were
then normalized to the continuum level using IRAF\footnote{IRAF is distributed
by the National Optical Astronomy Observatory, which is operated by the
Association of Universities for Research in Astronomy (AURA) under a cooperative
agreement with the National Science Foundation.}, and the same normalization was
applied to the Stokes V and null N spectra.

\begin{figure}
\resizebox{\hsize}{!}{\includegraphics[clip,angle=-90]{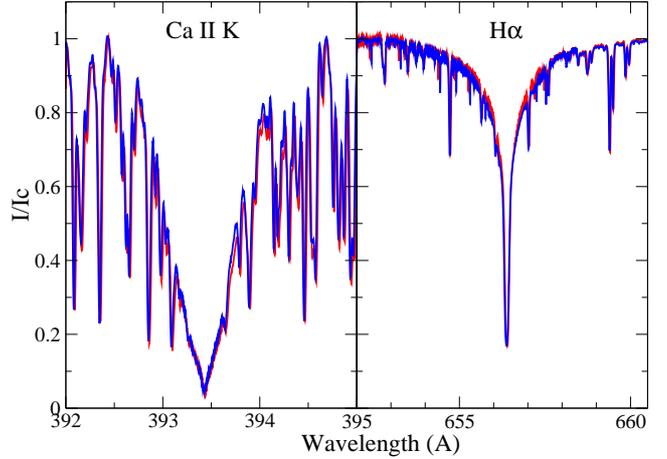}}
\caption[]{\ion{Ca}{ii} K and H$\alpha$ lines for the first
(red line) and second (blue line) average spectra of
$\rho$\,Pup.}
\label{CaHalines}
\end{figure}

$\rho$\,Pup is known to host a chromosphere, detectable as emission in the UV
\ion{Mg}{ii} h and k lines \citep{fracassini1983}. Weak emission was also
claimed in the \ion{Ca}{ii} K line by \cite{dravins1977} but remained
unconfirmed by subsequent observations \citep{mathias1997,mathias1999}, possibly
indicating a transient nature due to a shock wave. In the ESPaDOnS spectra
presented here, we detect no emission, neither in the \ion{Ca}{ii} K line nor in
the H$\alpha$ line (see Fig.~\ref{CaHalines}).

\subsection{Magnetic field detection}

To check for the presence of a magnetic field, we used the Least Squares
Deconvolution (LSD) technique \citep{donati1997}. LSD requires a mask listing
the lines in the spectrum, their wavelength, depth, and Land\'e factor. To
produce this line mask, we first extracted a line list from the VALD3 atomic
database \citep{piskunov1995,kupka1999,vald3} for the typical values $T_{\rm
eff}=6750$ K and $\log g=3.5$ of $\rho$\,Pup. We only used lines with a depth
larger than 10\% of the continuum. We then removed from the mask all lines that
are not visible in the intensity spectra, hydrogen lines because of their
Lorentzian broadening, the lines blended with H lines or interstellar lines, as
well as lines in regions affected by telluric absorption. Finally, the depth of
each line in the LSD mask was adjusted to fit the observed line depth. The final
mask tailored for $\rho$\,Pup contains 12515 lines and was used to produce the
LSD Stokes profiles.

Consecutive sequences of spectra were co-added to produce one single magnetic
measurement per night. Our LSD procedure forces the LSD mean intensity
weight and mean polarisation weight to be equal to 1. As a consequence, the
mean Land\'e factor is 1.187 for the first observation and 1.188 for the second
observation. The mean wavelength is 548.94 nm for the first observation and
541.08 nm for the second observation. Adopting the standard wavelength of
500 nm, this corresponds to an equivalent Land\'e factor of 1.383 for the
first observation and 1.377 for the second observation.

\begin{figure}
\resizebox{\hsize}{!}{\includegraphics[clip,angle=-90]{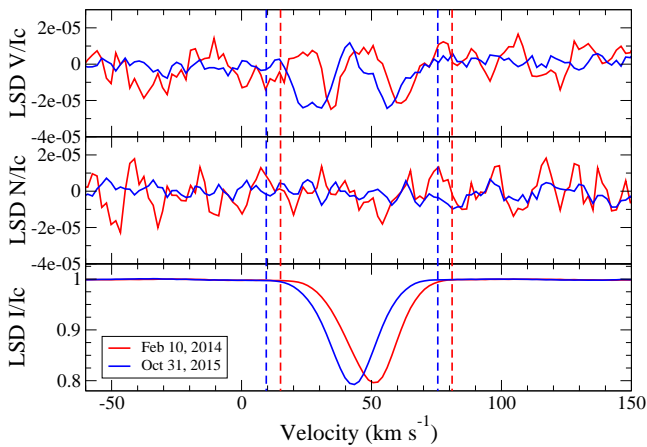}}
\caption[]{LSD Stokes V (top), N (middle) and I (bottom) profiles for the first
(red line) and second (blue line) spectropolarimetric measurements of
$\rho$\,Pup. Vertical dashed lines show the integration range for the $B_l$,
$N_l$ and FAP calculations for each measurement.}
\label{LSDprof}
\end{figure}

We find that both observations show a Zeeman signature in their Stokes V
profile, indicating that the star is magnetic (see Fig.~\ref{LSDprof}). The two
signatures, obtained in 2014 and 2015, are very similar in shape and shifted
with the intensity profile. As expected, thanks to the short duration of the
polarimetric sequence, the N profiles are flat and show only noise, i.e. the
measurements have not been polluted by the stellar pulsations. 

The formal statistical detection of a magnetic field is evaluated by the False
Alarm Probability (FAP) of the signature in the LSD Stokes V profile inside the
LSD line, compared to the mean noise level in the LSD Stokes V profile outside
the line. We adopted the convention defined by \cite{donati1997}: if the FAP is
below  0.001\% the magnetic detection is definite, if the FAP is between 0.001\%
and  0.1\% the detection is marginal, otherwise there is no magnetic detection.
The same procedure was also applied to the N profiles. 

We obtain no formal detection in the first observation, which has a low S/N.
Nevertheless, our experience is that the FAP criterion is relatively
conservative. We judge the signature in the line profile obtained on February
10, 2014, to be significant, and it is corroborated by our subsequent definite
detection of the field in the second observation from October 31, 2015, thanks
to its higher S/N. When applied to the N profiles instead of the V profiles, we
get no detection in both measurements (see Table~\ref{tableB}). 

\subsection{Magnetic field measurements}

From the Stokes V and I profiles of each night of observations, we can calculate
the longitudinal magnetic field value, i.e. the strength of the magnetic field
in the direction of the observer. We use the center-of-gravity method
\citep{rees1979,wade2000} over a velocity range of $\pm$33 km~s$^{-1}$ around
the line centroid (48 and 42.5 km~s$^{-1}$, respectively). Results are shown in
Table~\ref{tableB}. The longitudinal magnetic field of $\rho$\,Pup is very weak,
below 1 G.

\section{Discussion}

HD\,188774 was the only confirmed magnetic $\delta$\,Scuti star so far
\citep{neinerlampens2015}. \cite{kurtz2008} and \cite{hubrig2016} claimed
that the Ap star HD\,21190 is magnetic and pulsating, however \cite{bagnulo2012}
showed that the star is probably not an Ap star and the magnetic detection 
is likely spurious (as are most of the few-$\sigma$ detections obtained
with FORS). \cite{alecianwade2013} also claimed a possible magnetic detection
in the $\delta$\,Sct star HD\,35929. However, this object is a Herbig star and
the spectropolarimetric results were unclear. Therefore, the clear detection of
a magnetic field in the well known $\delta$\,Scuti star $\rho$\,Pup makes it the
second confirmed magnetic $\delta$\,Scuti star, after HD\,188774.

The magnetic field measured in $\rho$\,Pup is very weak. The weakness of the
surface field can only be partly explained by the evolutionary state of
$\rho$\,Pup. The current radius of $\rho$\,Pup is 3.52 R$_\odot$. Assuming
magnetic flux conservation, the surface field when the star was on the MS was
only about 5 times stronger than the current field.

The magnetic field inferred directly from the Stokes V profile corresponds only
to the line of sight component. Thus it is possible that $\rho$\,Pup possesses a
relatively strong dipole magnetic field that was viewed near the magnetic
equator (at which $B_l$ is null) in both observations. This is qualitatively
consistent with the ``crossover'' morphology of the Stokes V signatures observed
in both measurements. However, given the very high precision of the
observations, it seems unlikely that the dipole field would be stronger than a
few tens of G. Another possible (although admittedly unlikely) explanation for
the similar signatures observed in both measurements is that $\rho$\,Pup
possesses a surface magnetic field with a dominant toroidal configuration.
Testing these various scenarios will require monitoring and modeling of the
phase variation of the Zeeman signature.

Very weak fields have been discovered in several Am stars
\citep{petit2011,blazere2016_alhena,blazere2016_am}. In these stars, the
longitudinal field strength is usually below 1 G, except for Alhena for which it
is a few gauss \citep{blazere2016_alhena}. \cite{yushchenko2015} showed that the
chemical peculiarities observed in the spectrum of $\rho$\,Pup are similar to Am
stars. Therefore, it is likely that $\rho$\,Pup is a cooler, evolved
counterpart of magnetic Am stars, as also proposed by \cite{kurtz1976} and
\cite{gray1989}. The magnetic field of Am stars is thought to be of fossil
origin. The simplicity and similarities in the Stokes V profiles of the two
spectropolarimetric measurements presented here would corroborate this
hypothesis. In the case of a dynamo field, we would expect the Zeeman signature
to have changed more drastically between the two epochs. 

Finally, \cite{nardetto2014} showed that $\rho$\,Pup does not fit their expected
atmospheric velocity gradient curve (their Fig. 9). They proposed that this may
be related to the low peak-to-peak amplitude of the radial velocity curve.
Following the discovery of a magnetic field in this star, we propose that
magnetic effects could explain this discrepancy. Indeed, the action of the
magnetic field could modify the velocity gradients. 

\begin{table}
\caption{Detection probability in the Stokes V and N
profiles, detection status, and longitudinal magnetic field measurements from
the V and N profiles with their error bars for the two spectropolarimetric
measurements of $\rho$\,Pup.}
\begin{tabular}{llllll}
\hline
\# & Prob$_V$ & Prob$_N$ & Detection & $B_l \pm \sigma B$ & $N_l \pm \sigma N$ \\
 & \% & \% &  & G & G \\
\hline
1 & 11.5 &  0.9 & No       & -0.05$\pm$0.72 & -0.31$\pm$0.72 \\
2 & 100  & 23.8 & Definite & -0.29$\pm$0.32 &  0.35$\pm$0.32 \\
\hline
\end{tabular}
\label{tableB}
\end{table}

\section{Conclusions}

$\rho$\,Pup is a new magnetic $\delta$\,Scuti star, and only the second
confirmed member of this class. Its rather simple and stable Zeeman signatures
point towards a likely fossil origin of its magnetic field. Its very weak field
strength and chemical peculiarities link it to the recently discovered family of
ultra-weakly magnetic Am stars. 

Following the discovery of a magnetic field in $\rho$\,Pup, a complete
spectropolarimetric follow-up of this star over its rotation period should be
performed to characterise its field strength and topology in detail. Then, this
star will be an excellent target for magneto-asteroseismology: its radial and
non-radial pulsations combined to the magnetic information will allow one to
tidely constraint its seismic models. Moreover, it would be interesting to check
how the magnetic field varies during the chromospheric cycle and if it is
correlated with a possible shock wave.

\section*{Acknowledgements}

This research has made use of the SIMBAD database operated at CDS, Strasbourg
(France), and of NASA's Astrophysics Data System (ADS). GAW acknowledges
Discovery Grant support from the Natural Sciences and Engineering Research
Council of Canada.


\bibliographystyle{mnras}
\bibliography{articles} 

% Don't change these lines
\bsp	% typesetting comment
\label{lastpage}
\end{document}